\documentclass[useAMS,usenatbib]{aastex}     
\usepackage{spr-astr-addons,rotating,graphicx,natbib,epsfig,multirow,rotfloat}
\usepackage{threeparttable}
\begin{document}

\title{\bf The broad-band spectrum of the Narrow-line Seyfert 1 NGC 4748: from UV to hard X-ray}

\author{A.A. Vasylenko}
\affil{Main Astronomical Observatory, National Academy of Sciences of Ukraine, Zabolotnogo 27, 03680, Kyiv, Ukraine}
\email{vasylenko\_a@mao.kiev.ua} 

accepted for publication in

\textit{Astrophysics and Space Science}\\


\begin{abstract}

We report the results obtained by a broad-band (0.5-500 keV) data analysis of narrow-line Seyfert 1 galaxy NGC 4748 observed with an XMM-Newton/PN, INTEGRAL/ISGRI and SWIFT/BAT telescopes. This galaxy has a soft X-ray excess that is typical for the class of narrow-line Seyfert 1. The question of the origin of soft excess in such objects is still unclear. 
We tested and compared two spectral models for the soft X-ray spectra based on the different physical scenarios.
The first one is based on the Done \& Nayakshin model of two-phase accretion disc in a vertical direction, which includes two reflection zones with different ionization levels. According to this model, we found that a highly ionized reflection has the value of ionization $\xi\sim3000$ $erg~s^{-1}~cm$ and is mostly responsible for the soft excess. This reflection becomes comparable with a low ionized one ($\xi\sim30$ $erg~s^{-1}~cm$) in moderate X-ray range. However, this model requires also an additional component at soft energies with kT$\sim$300 eV. The second model is an energetically self-consistent model and assumes that a soft excess arises from optically thick thermal Comptonization of the disc emission. Combination of the UV (from XMM/Optical monitor) and X-ray data in the latter model allowed us to determine a mass of the central black hole of $6.9\times10^6 M_{\odot}$ and Eddington ratio $log_{L/L_{Edd}}\simeq-0.57$.
Also, we were not able to rule out one of competing models using only X-ray spectra of NGC 4748.

\end{abstract}


%

\section{Introduction}

The Narrow-Line Seyfert 1 (NLS1) as a specific peculiar subclass of galaxies with active nuclei (AGN) was defined by \cite{OP}. The key differences between NLS1 galaxies and ``normal'' Seyfert 1 (S1) galaxies are the smaller widths of H$_\beta$ line and weaker [OIII] emission \citep{BG}.

There are three common optical criteria of the emission line properties to ascertain whether AGN belongs to NLS1 subclass \citep{OP,G,BG,VVG}:
\begin{itemize}
	\item{the FWHM of the H$_\beta$ line is less than 2000 km/s;}
	\item{the ratio [OIII] 5007\AA/H$_\beta \leq$3};
	\item{strong FeII lines, i.e. the ratio R$_{4570}$  between fluxes or equivalent widths of FeII 4570\AA\ line and the broad H$\beta$ component R$_{4570}$, is $\geq$0.5.}
\end{itemize}

However, the first criterion is not sufficiently precise, because the value of 2000 km/s is some simplification of dividing \citep{G}. According to \citet[and references therein]{S,S1} and \cite{VZE}, the more flexible selection limit for NLS1 is the broad line width (FWHM) of H$_\beta$ $\sim$4000 km/s.

There are also a number of additional features in X-rays as an extension of the common criteria:

\begin{itemize}
	\item {prominent soft X-ray excess \cite[e.g.][]{Bol, VZE};}
	\item {steep X-ray power law emission (it is steeper than for broad-line S1s (BLS1)) \cite[e.g.][]{NP,VZE};}
	\item {X-ray variability \cite[e.g.][]{Bol, Dew, VZE}.}
\end{itemize}

Additionally, \cite{Z} have found a few objects with very faint or non-detectable Fe II emission line, which they called FeII-deficient NLS1 galaxies.

Recently the central black hole mass and Eddington ratio were estimated by \cite{CHE} for the sample of NLS1s taken from the 6dF Galaxy Survey. They obtained an average value of $9.55\times10^6M_{\odot}$ and $0.9L_{Edd}$ respectively, which is a typical value for NLS1s. As for the influence of large-scale environment on the intrinsic properties, the NLS1 host galaxies reside in less dense environments and their distribution is different as compare with BLS1 galaxies; moreover, a large fraction of NLS1 can be classified as having pseudobulges that is in favor of secular processes in the past evolution of their hosts rather than merging processes \citep{MAT, Pul, JAV}. It could confirm their younger age, and that BLS1s could be parent population of NLS1s and unified by orientation.

The first detailed X-ray spectral analysis of NLS1 galaxies has been presented in \cite{L1999} for 23 objects using ASCA and ROSAT data. Later \cite{BGM} performed the investigation of the properties of 21 NLS1 based on XMM-Newton data and showed that the soft/hard X-ray luminosity ratio of a NLS1 galaxy is typically higher than 1. This can be considered as a sign of the dominance of the soft component in the X-ray spectra. The origin of this soft excess is still unclear \cite[e.g.][]{GD, SD}.

According to \cite{CE}, the soft excess in NLS1s could be interpreted as a high energy tail of the thermal emission generated in the innermost regions of the accretion disc. Recently, an analysis of the UV/X-ray spectrum of II Zw 177 and its variability have been performed by \cite{P} using two XMM-Newton observations. They found that both blurred reflection model from an ionized disc and intrinsic disc Comptonization model describe soft excess well. In the same time, \cite{J2} have demonstrated that the soft X-ray spectra of NLS1 RX J1140.1+0307 can be reasonably fitted by blurred ionized reflection as well as optically thick Comptonization models too. These authors also pointed out that the latter model is more preferable.

The presence of these X-ray spectral features in NLS1 could be caused by the relatively small mass of the black hole ($\sim 10^6M_{\odot}$) with near-Eddingtonian (or even super-Eddingtonian) mass accretion rate \citep[e.g.][]{PDO, Boro, Grupe, M2, Z}. This means that the soft X-ray NLS1 spectrum is emitted mainly from the accretion disc (Jin et al. 2012; Done et al. 2012). Thus using the optical/UV disc normalization from the wide Spectral Energy Distribution (SED) one can determine black hole mass, accretion rate, and some parameters of the innermost part of the accretion disc (see e.g. \cite{D,J}). This is an independent method (e.g., the reverberation mapping) to determine the black hole masses.

On the whole, the soft and middle X-ray spectra (i.e., from ~0.3 keV up to ~20 keV) of NLS1s mainly modelled in the frame of two scenarios:
\begin{itemize}
	\item{the power-law emission reflected by ionized, often relativistically smeared/blurred medium, \citep{C, F, F1, Z1, F2}.}
	\item{optically thick ($\tau\approx10$), cool ($kT\approx$0.1-0.3 keV) Comptonized thermal emission of an accretion disc \citep{MH, M, Dew7, Jin};}
\end{itemize}

The spectral properties of NLS1 galaxies at hard X-rays (above 10 keV) are not fully systematized yet. Suzaku observation of NLS1 galaxy SWIFT J2127.4+5654 in 0.5-50 keV have been analyzed by \cite{M3}. They found that central black hole in this AGN appears to be a Kerr black hole with an intermediate spin of $\sim0.6$. Using Suzaku data in 0.4-40 keV range, \cite{T} have analyzed the X-ray spectrum of NGC 4051. They showed that this X-ray source was highly variable on short timescales and a spectral model must include a complex partial-covering absorption to explain the shape of spectrum.
Recently, \cite{MATT} have presented analysis of the combined Swift/XRT and NuSTAR X-ray data of NLS1 NGC 5506 in 0.8-79 keV. They constrained a highest lower limit to the cut-off energy, which was obtained so far in an NLS1s by the NuSTAR as $\sim$350 keV ($E_{cut}=720_{-190}^{+130}$ keV). This limit may indicate an inefficient cooling of the corona due to low-energy photon flux.

As for the averaged X-ray spectral properties of samples of NLS1 galaxies, several works could be noted. The first analysis of the hard-X-ray properties of the sample of five NLS1 galaxies (KRL2007 163, KRL2007 385, WKK 6471, LEDA 835095, and LEDA 090334) from the third IBIS catalogue \citep{B} has been presented by \cite{M1}. Their processing of the Swift/XRT and INTEGRAL/IBIS data (up to 100 keV) has shown that the observed spectra of these five objects have an exponential high-energy cut-off below 60 keV and a very steep high value of photon index ($\Gamma=2.6 \pm 0.3$) in 20-100 keV energy band.

Later \cite{RWCP} and \cite{PdRB} analyzed partly overlapped samples of 14 NLS1 galaxies detected by INTEGRAL/IBIS in broad-band X-ray range. In both papers there were also obtained steep values of hard photon index $\langle\Gamma\rangle$ $\sim 2.3$.

Our study is focused on the NGC 4748, which has not yet been studied in detail in soft and medium X-ray range.

NGC 4748, which is known as MCG -02-33-034 or RX J1252.2-1324, with a redshift $z = 0.01463$ \citep{MR} is a quite nearby NLS1 galaxy in Corvus constellation.
It is a barred spiral galaxy interacting with a slightly smaller spiral galaxy  \citep{OdR, VV}. NGC 4748 is characterized by a radio-quiet active nucleus (1.4 Ghz flux of 14.0$\pm$0.6 mJy, \cite{Gal}) as well as a sub-nuclear starburst activity. The Fe II $\lambda$4570\AA/$H_\beta$ ratio of the object is equal to 0.55, and a H$_\beta$ line width of 1565 km/s \citep{VV}. The mass of the central black hole has been estimated to be $5.5 \times 10^6 M_{\odot}$ derived from the optical continuum luminosity \citep{Hao} (i.e. using the luminosity at 5100\AA\ and the FWHM of $H_\beta$ \citep{Ka}); $4.2 \times 10^7 M_{\odot}$, through the correlation of the BH mass and the velocity dispersion within the narrow line region indicated by the [OIII] line width \citep{WL}, and $2.55^{+0.74}_{-0.88} \times 10^6 M_{\odot}$ through the reverberation mapping method with HST/WFC3 data \citep{GMW}.

NGC 4748 has a quite steep X-ray photon-index $\Gamma = 2.50 \pm 0.17$ according to the ROSAT/HRI data \citep{PBR} and $\Gamma =  2.20 \pm 0.11$ according to the unabsorbed power-law model \citep{L} based on Swit/XRT spectrum. \cite{PdRB} obtained a photon index of $\Gamma = 2.01 \pm 0.13$ and no cut-off at high energies. No Fe-K emission lines and reflection components in the composite Swift/XRT + INTEGRAL/ISGRI spectrum were observed. They also found the luminosity ratio between 0.1-2.0 keV and 2-10 keV band slightly higher than 1.

In this paper we present our results on the detailed case-study of the NGC 4748 using the X-ray observational data obtained by INTEGRAL/ISGRI, XMM-Newton, and Swift/BAT missions. This allowed us to analyze a wide-band X-ray spectrum up to $\sim$500 keV.
 
The paper is organized as follows: Section 2 describes the observational data and their reduction; Section 3 contains a short timing analysis; Section 4 presents the results of spectral fitting, and Section 5 contains the discussion of the obtained results.

\section{Observations and data reduction}

We used the XMM-Newton (EPIC and OM), SWIFT/BAT and INTEGRAL/ISGRI datasets in our analysis. We describe below the reduction applied to the initial data to obtain the spectra and lightcurves.

\subsection{XMM-Newton}

NGC 4748 was observed by the XMM-Newton mission on 2014 January 14 (OBSID 0723100401, PI. B.Kelly) for $\sim$ 69 ks. During the observational time all three EPIC cameras were operated in Large-window mode using the Medium filters for MOS cameras and Thin1 filter for PN camera. The EPIC data were processed using the standard software package XMM SAS ver. 14.0. Only the data from EPIC PN camera were used due to their higher quantum efficiency and larger effective area at lower energies compared with the MOS cameras. The standard SAS chain \textit{epproc} was applied for primary data reduction. The single- and double-photon events were taken into account (i.e., the PATTERN$\leq$4). To exclude bad pixels and near-CCD-edge events from our consideration, the filter FLAG=0 was also applied. The source spectra and light curves were extracted from the source-centered 32 $\sec$-radii circular region, while the empty regions on the same CCD chip were chosen to extract the background counts. The periods of time with comparably high flaring particle background were removed from the cleaned event files using the \textit{tabgtigen} task. Pile-up check, performed with the \textit{epatplot} metatask had detected the presence of a slight pile-up in the PN image. That is why we have excluded the central region of the source with the radius of 10 $\sec$ from our consideration to fix this problem. Unfortunately, we could not corrected completely the pile-up effect for double events. By this reason, we worked only with single events (i.e., PATTERN==0).

Ancillary files and response matrices were obtained using the standard chains \textit{arfgen} and \textit{rmfgen}. The extraction of background-subtracted lightcurves was made by the standard \textit{lcmath} task included in FTOOLS. The effective exposure of the PN observation is 28190 s, with the averaged count rate on the level of 2.99 cts/s. Finally, the spectrum in 0.5-10 keV range was rebinned with at least 30 counts in a spectral channel.

Within this XMM-Newton observation of NGC 4748, the UV observations by the Optical monitor (OM) in UVM2, UVW1, and U filters were also available. These data were processed using the specialized SAS metatask \textit{omichain}. Then, \textit{om2pha} chain was performed to obtain the OGIP II-type file for simultaneous spectral fitting with the EPIC data as additional spectral bins. We have added 5\% systematic errors to these UV bins. We also corrected our UV data for the Galactic extinction following \cite{SF} and \cite{SFD}, by E(B-V)=0.0441.

\subsection{INTEGRAL \& Swift}

The {\it ISGRI} dataset of the {\it INTEGRAL} observations analyzed in the given paper includes all the data publicly available in {\it INTEGRAL} data archive on 1st April 2015, i.e. spacecraft revolutions from 0077 to 1080. The total ISGRI exposure time of the dataset, which had been used, is 1.04 Ms (including all the observations when the object was at the angle less than $10^\circ$ off-axis).

We performed {\it INTEGRAL} IBIS/ISGRI data analy\-sis with version 10.1 of the Off-line Scientific Analysis software (OSA). We used standard recipes of spectral extraction for IBIS/ISGRI and OSA software. All the spectra were extracted individually for every science window and then summed up into the whole set. The source is observed by ISGRI up to $\sim 500\,keV$. The significances of detection are $29\sigma$ in $17 \div 80 \,keV$ and 30$\sigma$ in $80 \div 250\, keV$.

The Swift/BAT stacked spectra in 14-195 keV ener\-gy range and Crab-weighted light curve are avai\-lable from the 70 month catalog (2004-2010)\footnote{http://swift.gsfc.nasa.gov/results/bs70mon/SWIFT\_J1252.3-1323}. The effective exposure time is 7.093 Ms, with the average count-rate $2.04 \times 10^{-5}$ cts/s.

\section{Lightcurves and variability}

Figure 1 shows the XMM-Newton background-sub\-tracted light curves of NGC 4748 obtained by EPIC PN camera in two bands - 0.5-2.0 keV and 2-10 keV and their ratio (2-10 keV/0.5-2.0 keV). During the XMM-Newton observation, the light curve show a significant variability, as it expected for NLS1 type of galaxies. To check the time variabilities during the observation, we applied the FTOOLS task \textit{lcstats} to each light curve. This task performs statistical analysis, and, among other values, provides the constant source probability, associated to the $\chi^2$ value and RMS fractional variation. Thus, assuming a hypothesis of a constant light curve, we obtained $\chi^2$/d.o.f. = 2529/65 ($f_{rms,soft}\sim 18$\%) in the soft band and $\chi^2$/d.o.f. = 199/65 ($f_{rms,hard}\sim 14$\%) in the hard band, which indicate a significant variability in both energy ranges. The average count rate  is 4.81$\pm$0.02 cts/s in the 0.5-2.0 keV band and 0.77$\pm$0.09 cts/s in the 2.0-10.0 keV. The detailed timing analysis of the observed light curves is not the aim of this paper, and presented in a separate paper \citep{Fed}. In this article, we performed the spectral analysis for the time-averaged spectrum.

\begin{figure}[h]
\centering
\epsfig{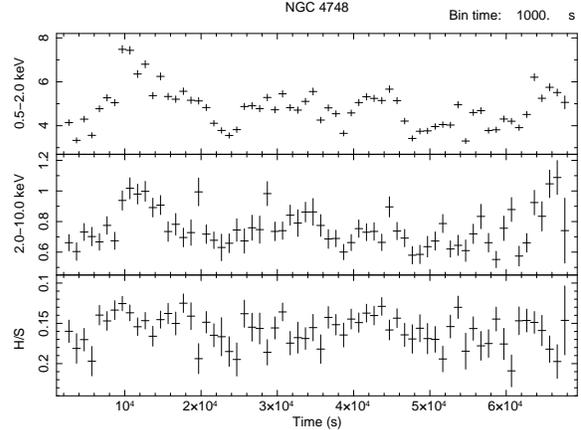}
\caption{\fontsize{8}{13}\selectfont XMM/EPIC lightcurves of NGC 4748 in 0.5-2.0 kev and 2-10 keV energy ranges and their ratio.}\label{Fig1}
\end{figure}

\section{Spectral fitting}

In our analysis we have applied several models of different levels of complexity, for the EPIC/PN spectrum alone or fitted together with OM, Swift/BAT and INTEGRAL/ISGRI spectra.

All the spectral fits were performed using XSpec v.12.8 of HEASOFT package, version 6.16. The Galactic absorption was modelled using {\tt tbabs} model by \cite{WAM} and included as the fixed value of $3.52 \times 10^{20} cm^{-2}$ in all the models, following the results by Leiden/Argentine/Bonn (LAB) Survey of Galactic HI \citep{K}\footnote{https://heasarc.gsfc.nasa.gov/cgi-bin/Tools/w3nh/w3nh.pl}. The inclination angle to the line-of-the-sight was frozen to the value of 52$^\circ$ following the information from LEDA database. The value of the intercalibration constant used in our simultaneous modelling of XMM-Newton/EPIC, INTEGRAL/ISGRI, Swift/BAT spectra was in the range 1.2-0.8, that means within the typical range of values 2.0-0.5. The errors, lower and upper limits correspond to a 90\% confidence level ($\Delta \chi^2=2.71$) for one interesting parameter. Throughout the paper we adopt the following standard cosmological parameters: $H_{0}$ = 70 km s$^{-1}$ Mpc$^{-1}$, $\Lambda_{0}$ = 0.73, $\Omega_{M}$ = 0.27.

\subsection{`Baseline' models}

We started our spectral analysis from a simple power-law model with galactic $N_\mathrm{H}$ applied 
to the EPIC/PN spectral data in 2-10 keV energy range. This model provides fit giving a photon index $\Gamma$=2.00$\pm$0.06 and a minimum $\chi^2$/d.o.f. = 163/141. If we extrapolate this model to the soft X-ray band (up to 0.5 keV), we obtain a very worse fit with $\chi^2$/d.o.f. = 6709/427, that means a presence of a significant soft excess in the spectrum. In order to model the prominent soft excess, we used the phenomenological {\tt zbremss} model, which is a single temperature redshifted thermal bremsstrahlung emission model. The addition of this component improves the fit significantly, with $\chi^2$/d.o.f. = 456/424. The temperature of the soft component is found to be $\sim297$ eV, which is moderately high value. We did not find a presence of any addition neutral/ionized absorber but the adding two edges at the energies of $0.73\pm$0.04 keV and $0.61\pm$0.03 keV with $0.20\pm$0.08 and $0.25\pm$0.13 optical depth, respectively, improves the fit of $\chi^2$/d.o.f. = 433/421 ($\Delta \chi^2=28$ for three additional d.o.f.). These features could be interpreted as O VI-VII edges (E$_{O VI\ edge}^{rest\ frame}$= 0.671 keV, E$_{O VII\ edge}^{rest\ frame}$= 0.739 keV) or N VII (E$_{N VII\ edge}^{rest\ frame}$= 0.667 keV) instead of O VI edge. In order to determine the statistical significance of the absorption edges (due to that the F-test does not give the correct result for the multiplicative absorption component - see appendix A in \cite{Orl}), we used the Xspec \textit{simftest} script with 5000 trials for each of edges and found that the probabilities for the absorption edges to be required are $\sim99.97\%$ (for edge at $\sim0.61$ keV) and $\sim99.9995\%$ (for edge at $\sim0.73$ keV).

The derived spectral model provides a good fit but a slight excess between 5-7 keV is still present. For this reason, we added one narrow Gaussian emission line with fixed line width at a value of $\sigma$ = 0.01 keV, thus improving the quality of fit to $\chi^2$/d.o.f. = 425/423. We found the centroid value of the line energy to be 6.43$_{-0.14}^{+0.11}$ keV (that is consistent with neutral Fe K$_{\alpha}$ line) and the equivalent width $EW = 115_{-61}^{+76}$ eV. However, after addition of a narrow iron line, there were still residuals presented at energies $\sim7$ keV. Therefore, we added a second line component to the model and refit the data. The fit gives a centroid energy of 7.03$_{-0.33}^{+0.19}$ keV and a line equivalent width of $EW = 245_{-124}^{+112}$ eV. This emission feature could be interpreted as emission line from Fe XXVI ions. The inclusion of this component leads to a $\chi^2$/d.o.f. = 419/420. However, inclusion of these two emission Fe lines changes the resulting statistic of $\Delta \chi^2=14$ for five additional d.o.f. So, if there are Fe lines in the spectrum\footnote{Albeit that the F-test shows a null hypothesis probability here of 0.025, this test is not statistically justified in the testing of presence of emission lines and can be used only for evaluation (see \cite{Prot}). Therefore, the Monte Carlo simulations were used in order to check the significance of improvement for the fits with Fe lines. Generating of 5000 fake spectra for each case give us the null-hypothesis probability of $2.9\%$ (i.e. without Fe K$_{\alpha}$) and $10.5\%$ (i.e. without Fe XXVI)}, they are weak ones.

Because the soft excess description by {\tt zbremss} model is phenomenological, we tried to replace it by a more physical Comptonized blackbody model {\tt compbb}. The fit with this model produces an acceptable fit of $\chi^2$/d.o.f. = 426/423 and returns the temperature of blackbody and optical depth of the plasma are $\sim150$ eV and $\tau \sim0.23$, respectively. This fit also gives an increasing value of the optical depth of edges by a factor $\sim3$ (see col. 3 in Table~\ref{tab1}).

In order to model a possible contribution of reflection spectral component and to take into account the presence of Fe lines, we also applied physical modeling. To begin we tried to replace the powerlaw and Fe Gaussian line by the {\tt pexmon} model \citep{N}, which includes the expected Fe K and Ni K emission lines self-consistently with the Compton reflection. This model provides an excellent fit ($\chi^2$/d.o.f. = 419/426) with best-fit values of photon index $\Gamma$=2.11$\pm$0.04, reflection parameter $R =0.83\pm$0.36 and slightly overabundance of $A_{Fe}=1.16_{-0.86}^{+1.76}$. The similar result of reflection in NLS1s was obtained, for example, by \cite{PdRB} and \cite{VZF}. However, {\tt pexmon} model includes only ``cold'' reflection. The ionization of the reflecting material was taken into account within the better reflection model {\tt xillver} \citep{GK,GDR}. Thereby, {\tt pexmon} model was replaced by more physically realistic XSTAR-generated {\tt xillver} model. In addition to the disc reflection, we include an intrinsic powerlaw continuum and for {\tt xillver} component we linking photon index to the one of the powerlaw component. The resulting fit has $\chi^2$/d.o.f. = 420/426. The best-fit values of parameters are $\Gamma$=2.09$\pm$0.04, ionization $\xi$=61$_{-27}^{+63}$ (quantified as $\xi= 4\pi F/n$ $erg~s^{-1}~cm$, where $F$ is the ionizing flux at the surface of the disc, and $n$ is the density). We note that if we assume the Fe abundance as a free parameter, we get the value $A_{Fe}\approx$13, but if we fix it $A_{Fe}\equiv$1, then the ionization value will change very slightly $\xi\sim54$.

The best-fit parameters for this ``baseline'' fitting of EPIC/PN spectrum are listed in Table~\ref{tab1}.

As a next step in our analysis, we included the INTEGRAL/ISGRI and Swift/BAT spectra, extending the energy range to the hard X-rays. We modelled these data using both of the reflection components - {\tt pexmon} and {\tt xillver}, which was described above. Both models proved to be a good fit: $\chi^2$/d.o.f. = 425/438 and $\chi^2$/d.o.f. = 426/434 obtained from PN+ISGRI data and $\chi^2$/d.o.f. = 427/434 and $\chi^2$/d.o.f. = 426/434 obtained from PN+BAT data respectively. Best-fitting parameters are presented in Table~\ref{tab2}.
We defined a reflection flux fraction by {\tt xillver} model fitting, as the ratio between unabsorbed 20-40 keV fluxes of the reflected and the incident components (as it is described by e.g., \cite{GKM} and \cite{Daus}). This fraction is about 0.5 for both PN+ISGRI and PN+BAT spectra.

\subsection{The broad-band X-ray double reflection model}

The physical explanation of the soft excess in galaxies such NLS1 is still debated. As it was already mentioned in Section 1, two possible explanations exist. According to first of them, the soft excess is a reflection from partially ionized accretion disc regions.

We tried to test a such scenario according to the assumption proposed by \cite{DN}, where the accretion disc has a discontinuous vertical structure with two skins that have a different ionization state. For simplicity, we do not include any relativistic smearing model (such as {\tt relxill} \citep{GDL}). We will analyze X-ray spectrum of NGC 4748 with this component in the separate paper.

Thus, we adopted a model, which consists of two {\tt xillver} (to account of both the neutral and ionized reflection) components and one power-law as the illuminating continuum. For the reflection component we tied both photon indices to the one of the power-law. The Fe abundances of two {\tt xillver} components are tied together too. The fitting of EPIC/PN plus INTEGRAL/ISGRI data with this model does not give a very good fit of $\chi^2$/d.o.f. = 476/438 with $\Gamma$=2.27$\pm$0.03 and the Fe abundance $A_{Fe} = 2.43_{-0.48}^{+0.45}$. The ionization parameters are found to be $\xi_{ion}$=1418$_{-388}^{+2341}$ and $\xi_{neutr}$=12$_{-2}^{+3}$. Since this variant of model most poorly agrees with the data at low energies, we decided to add the thermal bremsstrahlung model {\tt zbremss} and one edge.
This modification improves fit significantly $\chi^2$/d.o.f. = 418/437.
The photon index of power-law component becomes now slightly harder $\Gamma$=2.07$\pm$0.05. The Fe abundance increases to $4.34_{-1.27}^{+3.13}$, the ionized parameters changes to $\xi_{ion}=3070_{-1234}^{+3293}$ and $\xi_{neutr}=37_{-17}^{+39}$. If we calculate reflection flux fraction as it shown in the previous subsection, we then obtain the following values $R_{F\,neutr}\sim0.29$ and $R_{F\,ion}\sim0.38$ in the 20-40 keV band. We also calculated these fractions in 0.5-2.0 keV band in order to determine the relative contribution of ionized reflection into the soft excess and found $R_{F\,ion}\sim0.6$. Along with it, the neutral reflection value is found to be significantly smaller $R_{F\,neutr}\sim0.04$.

We also apply this model to the EPIC/PN plus Swift/BAT spectra. This gives us an excellent fit of $\chi^2$/d.o.f. = 426/431. The best-fit parameters are $\Gamma$=2.10$\pm$0.05, $A_{Fe} = 4.46_{-0.94}^{+1.28}$, $\xi_{neutr}=30_{-18}^{+30}$ and $\xi_{ion}=2731_{-410}^{+988}$. The calculated values of reflection fraction change slightly. In particular, $R_{F\,neutr}\sim0.03$, $R_{F\,ion}\sim0.42$ in the 0.5-2.0 keV band and $R_{F\,neutr}\sim0.39$, $R_{F\,ion}\sim0.30$ in the 20-40 keV band.

\begin{figure}[h!]
 \centering
    \epsfig{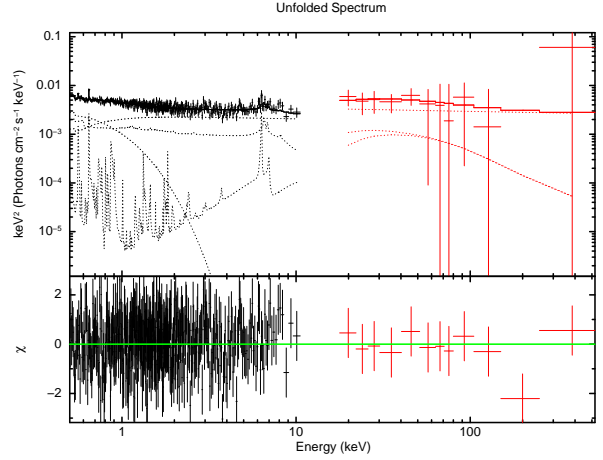}
    \caption{\fontsize{8}{13}\selectfont Unfolded 0.5-12 keV EPIC PN spectrum (black) with INTEGRAL/ISGRI 20-500 keV data (red) and the best-fitting ionized+neutral reflection model.} \label{Fig2}
\end{figure}
\begin{figure}[h!]
\centering
    \epsfig{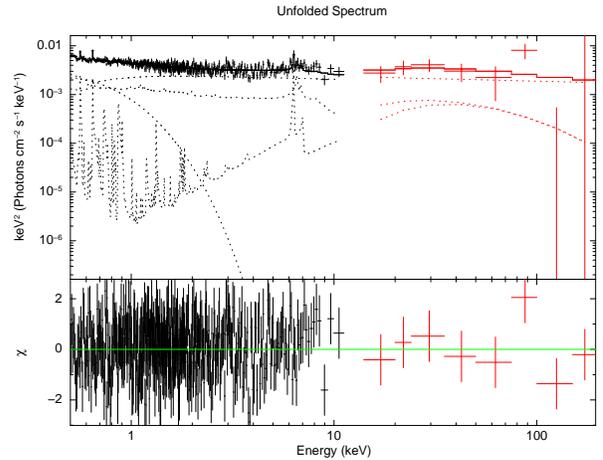}
    \caption{\fontsize{8}{13}\selectfont Unfolded 0.5-12 keV EPIC PN spectrum (black) with Swift/BAT 14-195 keV data (red) and the best-fitting ionized+neutral reflection model.} \label{Fig3}
\end{figure}

For completeness, we fitted simultaneously all three datasets (PN, BAT and ISGRI). The values of spectral parameters returned from this fit are in general compatible with those found from the preceding ones of pairs of spectra. The fit gave a $\chi^2$/d.o.f = 429/443. We again obtained reflection values $R_{F\,neutr}\sim0.04$, $R_{F\,ion}\sim0.59$ in the 0.5-2.0 keV band and $R_{F\,neutr}\sim0.26$, $R_{F\,ion}\sim0.35$ in the 20-40 keV band.

\begin{figure}[h!]
\centering
    \epsfig{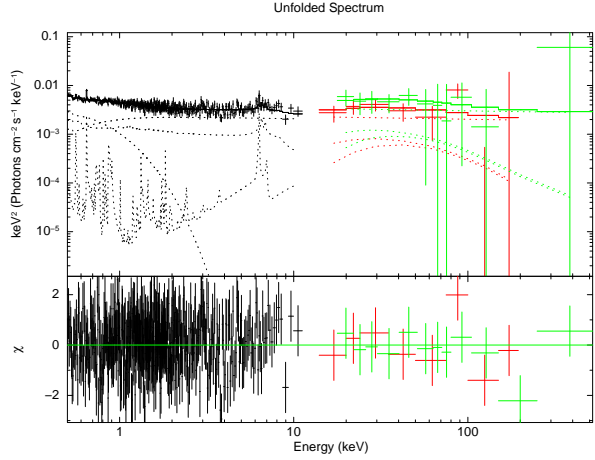}
     \caption{\fontsize{8}{13}\selectfont Unfolded spectra for combined data with the best-fitting ionized+neutral reflection model. Black refers to XMM-Newton/PN data, red to Swift/BAT and green to INTEGRAL/ISGRI}\label{Fig4}
\end{figure}

The observed data, the models, and the residuals are shown in Fig.~\ref{Fig2}, Fig.~\ref{Fig3} and Fig.~\ref{Fig4}, and the best-fit parameters for the double-reflection model are listed in Table~\ref{tab4}.

In summary of this subsection, we found that a joint fit of broad-band EPIC/PN plus ISGRI or/plus BAT spectra are described well by the model with reflection from two ionization states, and trend of changes of reflection fractions with energy is the same through all three combinations of spectra.

We will discuss our results in the Section 5.

\subsection{The UV to X-ray Comptonisation model}

The strong soft X-ray excess emission, except a reflection scenario, can also be described by opti\-cally thick  ($\tau\sim10$ and even higher) thermal Comptonization (with kT$_e$ $\sim0.1-0.2$ keV) in the inner part of accretion disc. In order to check this assumption, we adopted the physical model {\tt optxagnf} \citep{D}, which describes the AGN spectrum within the range from op\-tical/UV emission to hard X-rays. This model realizes the scenario including the intrinsic Compton emission upscattered in the warm optically thick medium of accretion disc, which responsible for the soft X-ray excess; power-law emission from optically thin hot corona as well as colour-corrected blackbody emission of thermalized accretion disc (UV band). The main interesting parameters of this model are as follows:
\begin{enumerate}
	\item $M_{BH}$, the central black hole mass;
	\item $L/L_{Edd}$, Eddington ratio;
	\item R$_{cor}$, the coronal radius, in units of the gravitational radius $R_g$;
	\item $kT_e$, the soft excess electron temperature;
	\item $f_{pl}$, the fraction of power between R$_{cor}$ and R$_{ISCO}$, that emitted as high-energy Comptonized component, charac\-te\-ri\-zed by the power-law model.  
\end{enumerate}

\begin{figure}[h!]
\centering
    \epsfig{file=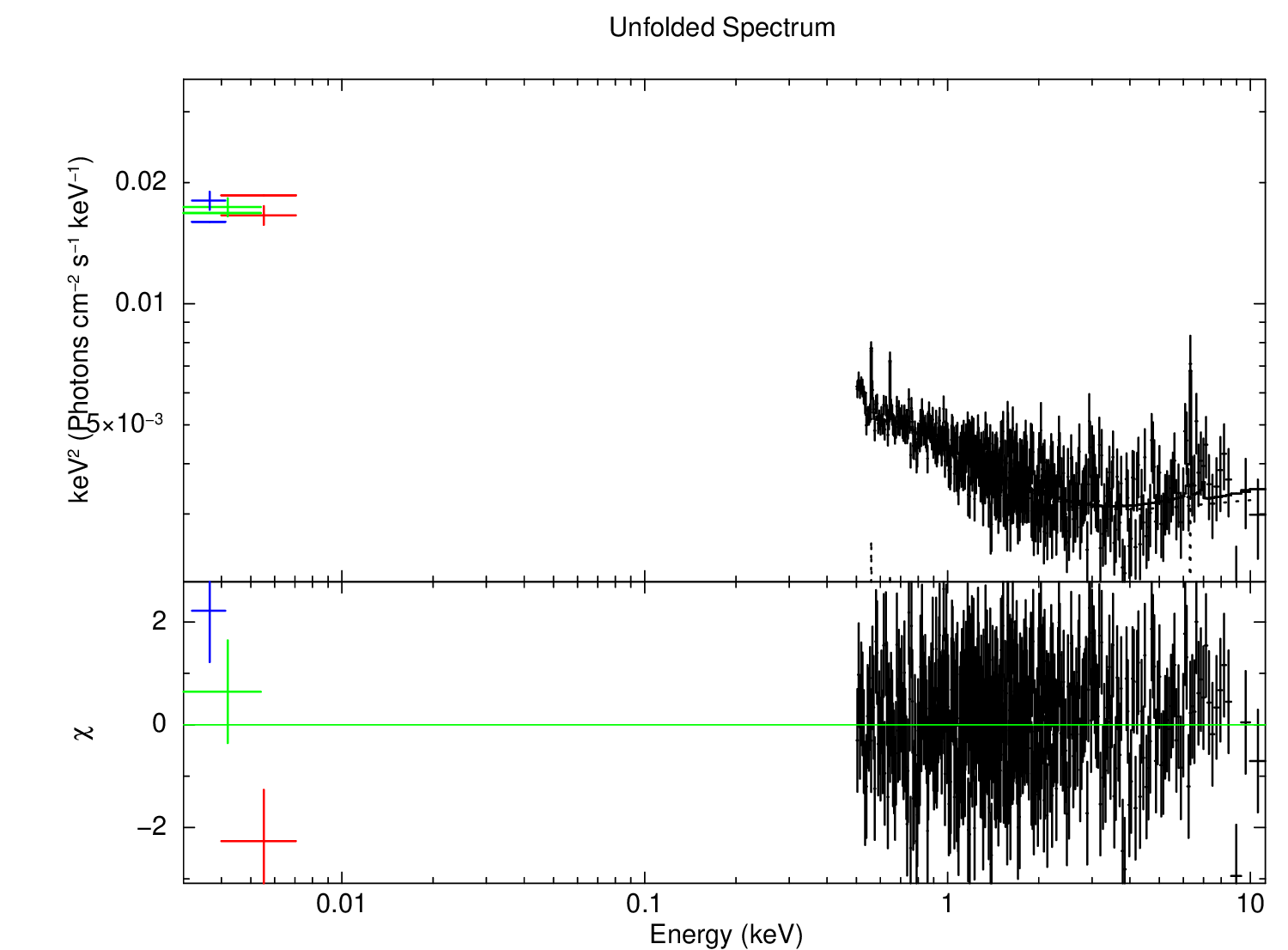,width=0.73\linewidth,angle=-90}
    \caption{\fontsize{8}{13}\selectfont Unfolded 0.5-12 keV EPIC PN spectrum (black) with UV Optical monitor data and the best-fitting intrinsic optxagnf model.} \label{Fig5}
\vskip 3mm
\centering
    \epsfig{file=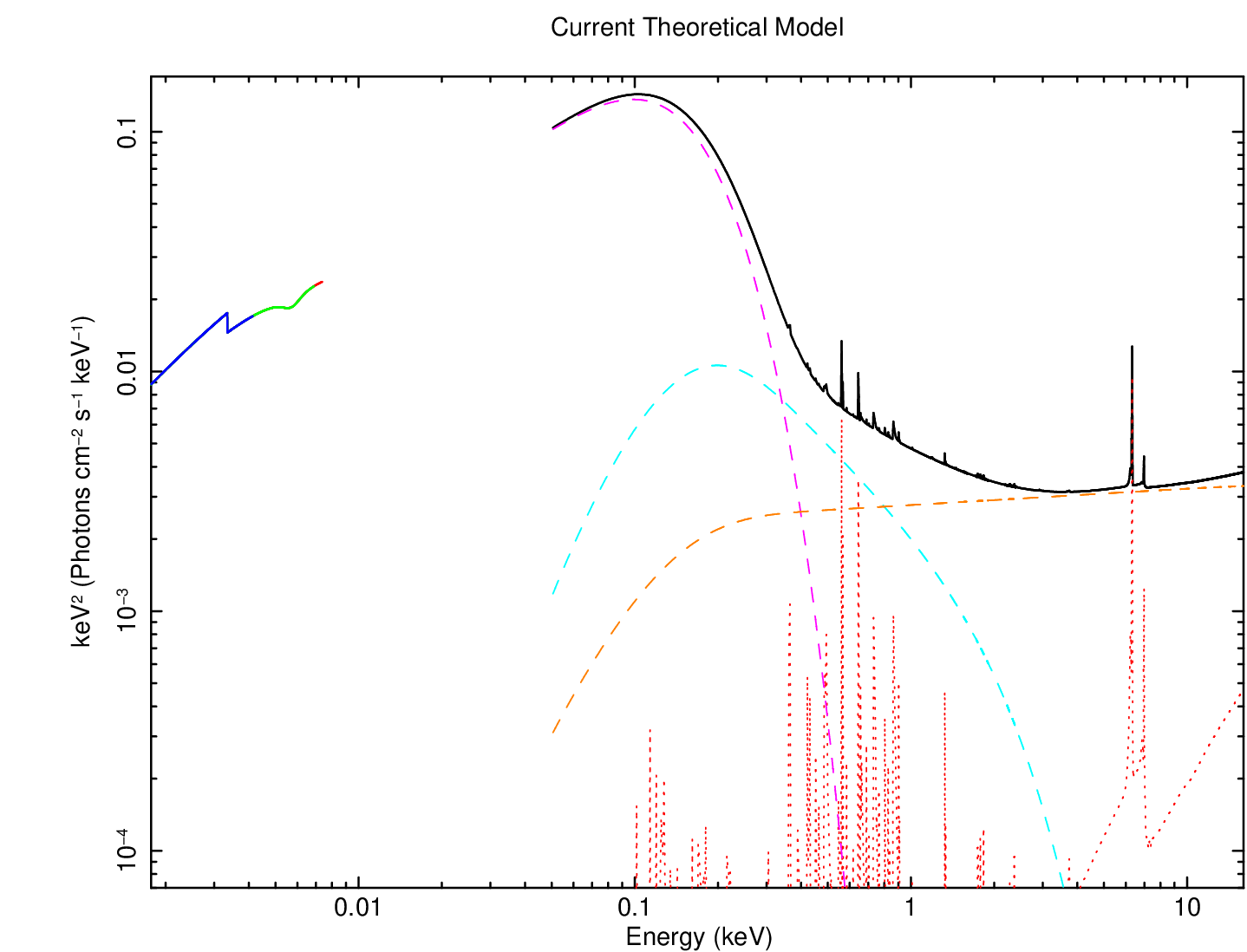,width=0.73\linewidth,angle=-90}
    \caption{\fontsize{8}{13}\selectfont The best-fit model of the UV+PN data with four individual components of the continuum. The red dotted line represents the {\tt xillver} model. The orange, cyan and magenta dashed lines represent the hard Compton, soft Compton and blackbody components of the {\tt optxagnf} model, respectively. The total spectra are plotted as the solid lines.} \label{Fig6}
\end{figure}

An additional model {\tt xillver} has been added to fit the reflection spectral features. A Galactic reddening factor has been modeled by {\tt uvred} model. We tried to fit the EPIC/PN and Optical monitor UV data together using
this model with a fixed value of the black hole mass at $2.55 \times 10^6 M_\odot$ by \cite{BK}; the spin value also was frozen to 0. The resulting fit, however, was not acceptable, yielding $\chi^2$/d.o.f. = 1137/540, which is mainly due to significant underestimation value of UV emission. Therefore, we tried to use the black hole mass as a free parameter and obtained significantly better fit of $\chi^2$/d.o.f. = 426/428 with $M_{BH}=(6.9 \pm 0.4) \times 10^6 M_\odot$ and coronal radius R$_{cor} = 11.3 \pm 0.5 R_g$. The best-fit values of other parameters are $log_{L/L_{Edd}}=-0.57\pm$0.02, $\Gamma$=1.94$\pm$0.01. One can see that the value of the black hole mass has been increased that consists with the results, obtained by other authors (see e.g. \cite{Jin, Star}). Also, the obtained value of the coronal radius is in agreement with the expected ones for NLS1 \citep{D, Ja, J2}. The observed data, the model, and the residuals are shown in Fig.~\ref{Fig5}, the best-fit model (without Galactic absorption {\tt tbabs} for clarity) is shown in Fig.~\ref{Fig6}, and the best-fit parameters are listed in Table~\ref{tab3}.

\section{Discussion}

\textit{The X-ray continuum and soft excess.} Our wide-band fitting allows us to compare two possible multicomponent models for description of the X-ray spectrum of the narrow-line Seyfert 1 galaxy NGC 4748.

The first model by \cite{DN} represents a sandwich-like disc consisting of layers with different density and ionization grades in the vertical dimension. These authors calculated the X-ray spectrum from the illuminated gas in hydrostatic balance. The soft excess in this model mainly originates as a result of reflection from the hot partially ionized skin of the accretion disc. 
According to this model, the behaviour of power-law index $\Gamma$ is different in relation to ionization parameter $\xi$ and optical depth of hot layer $\tau_{h}$. Mainly, there is a direct correlation between $\Gamma$ and $\xi$, while dependence between $\Gamma$ and $\tau_{h}$ is inverse. Reflection features from low or mildly ionized material came from deeper layers of accretion disc under the ionized skin.

One can see from of our modelling that the fraction of highly ionized component within this model is significantly higher than of the mildly ionized one in 0.5-2.0 keV energy band. Namely, the mean values of reflection fraction are $\langle R_{F\,neutr}\rangle\sim0.04$, $\langle R_{F\,ion}\rangle\sim0.54$, while these fractions become nearly equal in 20-40 keV energy band ($\langle R_{F\,neutr}\rangle\sim0.31$, $\langle R_{F\,ion}\rangle\sim0.34$). This demonstrates that the ionized reflection component plays an essential role in the formation of a soft excess.

However, observational data also required an additional soft component, besides ionized reflection. This component was fitted by {\tt zbremss} model, which can be explained in two ways. Firstly, the soft component is really represented by bremsstrahlung process that occurs in the hot ionized skin. Secondly, it can be moderate low-temperature Comptonization that occurs in the inner disc parts.

The second (i.e. the {\tt optxagnf}) model is more focused on the soft excess and interprets this feature as a sum of the thermal emission of optically thick Comptonized disc while high-energy tail is generated by ionized, optically thin corona. Values of spectral parameters obtained within this model are physically reasonable, for example, the coronal radius (R$_{cor} = 11.3 \pm 0.5 R_g$), fraction of power $f_{pl}=0.53\pm0.01$, and Eddington ratio $log_{L/L_{Edd}}=-0.57\pm0.02$, which are in a good agreement with the scheme by \cite{D} and comparable with the values that were obtained by other authors for NLS1 galaxies (see e.g. \cite{Ja, J2, Lah}).

\textit{Fe overabundance.} The reflection models used for fitting have showed high values of iron abundance. However, there are no the very strong Fe lines, neither neutral nor ionized, in the X-ray spectrum of NGC 4748. At first glance, we meet here an obvious contradiction: a high value of iron abundance (obtained from the modelling of the continuum emission) is not accompanied by the prominent iron emission lines in the 6.0-7.0 keV energy range.

However, there are some possible explanations of the presence of significant iron amount. 
 
The first one is the effect of a radiative levitation described by \cite{Rey}. The point is that a radiation pressure of the strong UV/EUV emission in central disc regions could produce a net upward force on the moderately ionized Fe atoms significantly stronger than the vertical gravity. Hydrogen and helium-like ions of iron play the minimal role in this process. Radiative levitation crowds out iron ions from the accretion disc volume to its photosphere causing the increasing abundance of iron here. Following \cite{Rey}, a radiative levitation can play a prominent role in AGNs with a comparably low mass of the central black hole and quite high accretion rate. Thus, it should be a some connection between iron overabundance and high values of the central BH spin \citep{Gallo, Garcia}.

Another explanation of iron overabundance in the absence of strong iron emission lines between 6 and 7 keV can be associated with the presence of an ionized wind or outflow. If, as in our case, we have a quite high inclination of the accretion disc to the line-of-sight ($\sim50^\circ$) (i.e., comparable to the inclination disc of galaxy)\footnote{Of course, it is not necessary that the inclination angle of accretion disc and galaxy disc have the same value. However, we consider that our approach is reasonable, because of the reflection features are tracks of the inclination of accretion disc, and in our case, we have not a strong Fe K$\alpha$ emission line with the distinct relativistic profile or high-quality x-ray data in 20-50 keV band.}, then the emission of central part of AGN on its way to the observer will pass through this outflow or wind emanating from the disc surface. In this case, the inhomogeneities or highly ionized blobs in such outflow can cause significant absorption in lines, e.g., resonant absorption lines of FeXIX K$\alpha$ and FeXXVI K$\alpha$ near 6.4 and 6.9 keV, respectively. The similar effect was detected, for instance, in NGC 4258 by \cite{YW}. Moreover, outflows may be responsible for the presence of moderate O VI, O VII edges as evidence of warm absorption.        

Recently, \cite{GFK} discussed that such Fe overabundance, as well as soft excess, can be explained by models of reflection spectra with illuminated atmosphere densities larger (i.e., $n_{e}>10^{17} cm^{-3}$) than the commonly assumed ($n_{e}=10^{15} cm^{-3}$). Authors suggested that first feature arises through the underestimation of the Fe line production at high densities, and enhanced emission at soft energies occurs due to the significant increase of free-free heating (i.e., bremsstrahlung) that provides raising gas temperature.

\textit{The Fe emission lines and neutral reflection.} We found a presence of two weak emission lines, which are, presumably, the lines of neutral ($E_{FeK}$=6.43 keV, $EW\sim115$ eV) and highly ionized Fe ($E_{FeXXVI}$=7.02 keV, $EW\sim235$ eV). It is widely assumed that the Fe lines are formed in a thin surface layer of the accretion disc and follow, among others, the ionization state of matter in this region \citep{F4}. Additionally, the line broadening $\sigma$ allows us to determine the approximate location of Fe lines emission region in the radial direction. Thus, if we calculate the velocity broadening of FeXXVI line, we obtained $V_{FWHM}\sim$ 24000 $km\,s^{-1}$, but this value for neutral Fe line is only about $\sim$ 1100 $km\,s^{-1}$. This means that the ionized Fe line originates in the inner parts of an accretion disc, while the neutral Fe K$\alpha$ line originates in the outer parts, or/and in a dusty torus with a small covering factor.
It is in consistency with an assumption that the accretion disc becomes more ionized with decreasing of the radius.

The characteristics of neutral reflection and weak Fe K line described in Subsection 4.1 are in agreement with both of models discussed above because they can be simply explained by the geometrical factor which is irradiation of the external cold parts by a primary emission at grazing small angles. It is worth noting that \cite{NKK} had shown in the modeling of reflection with hydrostatic equilibrium that the incident X-ray photons when penetrating into the disc at the decreasing angle of reflective material will undergo a large number of scattering. As a result, a significant signature of the reflection spectrum will be absent.

The similar results has obtained by \cite{MERL}, where the accretion disc model with radiation pressure domination was discussed. This model assumed that inhomogeneous accretion flow in the inner regions has a multiphase structure in sub-Eddington regime that take place in NLS1 galaxies. The authors found that the contribution of reflection components will strongly depend on the number of small-scale cold clouds in a hot plasma. The amount of this cold matter may be described by using an effective `cloud' optical depth $\tau_{\beta}$\footnote{$\tau_{\beta}=\pi N \epsilon$, where N is a total number of clouds and $\epsilon$ is a size of cloud (see Section 2 in \cite{MC02})}. In general, the reflection parameter increases with increasing of the optical thickness. In our case, we have reflection $R \sim$ 0.8 and photon index $\Gamma \sim$ 2, obtained using the {\tt pexmon} model. Therefore, according to the model of inhomogeneous accretion flow, cold phase optical depth should have a low value $\tau_{\beta} \sim$ 1-2 (see Fig. 2 in \cite{MERL}). In addition such value of $\tau_{\beta}$ provides a clear variability of the X-ray light curve. However, we must be careful with the interpretation, because shape of spectra and light curve characteristics also depend strongly on the geometry and relative position of hot and cold regions.

\section{Conclusions}

In this paper, we have presented an analysis of the time-averaged broad-band X-ray spectrum of NGC 4748. We examined two spectral models - the double reflection and the intrinsic disc Comptonization. Both of spectral models provide us physically reasonable parameters, even despite the obviously exaggerated value of iron abundance.

The first of these models represents the hypothesis of a two-component accretion disc, namely a strongly ionized inner disc and a mildly ionized outer disc. From this approach, we found that contribution of reflection with respect to the power-law continuum is changeable with ionization state and energy range, i.e. $\langle R_{F\,neutr}\rangle\sim0.04$ and $\langle R_{F\,ion}\rangle\sim0.54$ in 0.5-2.0 keV band and $\langle R_{F\,neutr}\rangle\sim0.31$ and $\langle R_{F\,ion}\rangle\sim0.34$ in 20-40 keV band. Consequently, the ionized reflection is responsible, at least, for a significant part of the soft excess, while it becomes comparable with the fraction value of the neutral reflection in the higher energy range.

The second model can reproduce the UV and X-ray data together and describes the soft excess as thermal Comptonization in the inner disc. As a result, an estimation of the central BH mass of $\sim7 \times 10^6 M_{\odot}$ and Eddington ratio of $L/L_{Edd}\sim0.3$ has been obtained.

However, these two models are not statistically distinguishable with the available observational data. Nevertheless, the nature of soft excess can be established via correlation variability between different X-ray spectral components in combination with another energy band. For example, this may be the variability between the X-ray soft excess and UV emission \citep{P} or between optical \& X-ray data (\cite{MEH} and \cite{CSV}). To understand more deeply the nature of NLS1 galaxies and NGC 4748 in particular, we preferably need the long simultaneous observations of XMM-Newton and NuSTAR which allow us, among others, to investigate the Compton hump energy range more precisely, that is very important for a model with two reflecting regions.

%
%

%
%

%

%
%

%

%

\section{Acknowledgements}

This paper is based on the observations obtained by XMM-Newton, an ESA science mission with instruments and contributions directly funded by ESA Member States and the USA (NASA). 
This work made use of data supplied by the UK Swift Science Data Centre at the University of Leicester, and data provided by the High Energy Astrophysics Science Archive Research Center 
(HEASARC), which is a service of the Astrophysics Science Division at NASA/GSFC and the High Energy Astrophysics Division of the Smithsonian Astrophysical Observatory. 
This research has made use of the NASA/IPAC Extragalactic Database (NED) which is operated by the Jet Propulsion Laboratory, California Institute of Technology, under contract with the NASA.

The author is grateful to Dr. Kryshtal O.N. and Dr. Fedorova E.V. for the useful discussion, Dr. Vavilova I.B. for helpful remarks and Dr. Chichuan Jin for his kind instructions on the correct use of {\tt optxagnf} model. The author also thanks Torbaniuk O. and Dr. Pakuliak L. for technical assistance.
The author is grateful to the anonymous referee for the helpful constructive comments and suggestions.

%

\begin{thebibliography}{99}
\fontsize{8}{11}\selectfont
\bibitem[\protect\citeauthoryear{Antonucci \& Miller}{1985}]{AM} Antonucci R., Miller J., 1985, \apj, \textbf{297}, 621.
\bibitem[\protect\citeauthoryear{Bentz \& Katz}{2015}]{BK} Bentz M. C., Katz S., 2015, \pasp, \textbf{127}, 67.
\bibitem[\protect\citeauthoryear{Bentz et al.}{2013}]{Bentz} Bentz M. C., Kelly D. D., Grier C. J. et al., 2013, \apj, \textbf{767}, 149.
\bibitem[\protect\citeauthoryear{Bianchi et al.}{2009}]{BGM} Bianchi S., Guainazzi M., Matt G., 2009, \aap, \textbf{495}, 421.
\bibitem[\protect\citeauthoryear{Bird et al.}{2007}]{B} Bird A. J. et al., 2007, \apjs, \textbf{170}, 175.
\bibitem[\protect\citeauthoryear{Bird et al.}{2010}]{Bird} Bird A. J. et al., 2010, \apjs, \textbf{186}, 1.
\bibitem[\protect\citeauthoryear{Boller}{1996}]{Bol} Boller T., Brandt W. N., Fink H., 1996, \aap, \textbf{305}, 53.
\bibitem[\protect\citeauthoryear{Boroson \& Green}{1992}]{BG} Boroson, T. A., Green, R. F., 1992, \apjs, \textbf{80}, 109.
\bibitem[\protect\citeauthoryear{Boroson}{2002}]{Boro}  Boroson T. A., 2002, \apj, \textbf{565}, 78.
\bibitem[\protect\citeauthoryear{Chen et al.}{2018}]{CHE} Chen S., Berton M., La Mura G., et al., 2018, \aap, in press (astro-ph/1801.07234).
\bibitem[\protect\citeauthoryear{Chesnok et al.}{2009}]{CSV} Chesnok N. G., Sergeev S. G., Vavilova I. B., 2009, KPCB, \textbf{25}, 107.
\bibitem[\protect\citeauthoryear{Comastri et al.}{1998}]{C} Comastri A. et al., 1998, \aap, \textbf{333}, 31.
\bibitem[\protect\citeauthoryear{Czerny \& Elvis}{1987}]{CE} Czerny B., Elvis M., 1987, \apj, \textbf{321}, 305.
\bibitem[\protect\citeauthoryear{Dauser et al.}{2016}]{Daus} Dauser T. et al., 2016, \aap, \textbf{590}, A76.
\bibitem[\protect\citeauthoryear{Dewangan et al.}{2002}]{Dew} Dewangan G.C. et al., 2002, \aap, \textbf{390}, 65D.
\bibitem[\protect\citeauthoryear{Dewangan et al.}{2007}]{Dew7} Dewangan G.C. et al., 2007, \apj, \textbf{671}, 1284.
\bibitem[\protect\citeauthoryear{Done et al.}{2012}]{D} Done C. et al., 2012, \mnras, \textbf{420}, 1848.
\bibitem[\protect\citeauthoryear{Done \& Nayakshin}{2001}]{DN} Done C., Nayakshin S., 2001, \apj, \textbf{546}, 419.
\bibitem[\protect\citeauthoryear{Fabian et al.}{2000}]{F4} Fabian A. C. et al., 2000, PASP, \textbf{112}, 1145.
\bibitem[\protect\citeauthoryear{Fabian et al.}{2002}]{F} Fabian A. C. et al., 2002, \mnras, \textbf{331}, L35.
\bibitem[\protect\citeauthoryear{Fabian et al.}{2004}]{F1} Fabian A. C. et al., 2004, \mnras, \textbf{353}, 1071.
\bibitem[\protect\citeauthoryear{Fabian}{2006}]{F0} Fabian A. C., 2006, Astron. Nachr., \textbf{327},943.
\bibitem[\protect\citeauthoryear{Fabian et al.}{2009}]{F3} Fabian A. C. et al., 2009, \nat, \textbf{459},540.
\bibitem[\protect\citeauthoryear{Fabian et al.}{2013}]{F2} Fabian A. C. et al., 2013, \mnras, \textbf{429}, 2917.
\bibitem[\protect\citeauthoryear{Fedorova}{2017}]{Fed} Fedorova E.V., 2017, BNUK, \textbf{56}, 18.
\bibitem[\protect\citeauthoryear{Gallimore et al.}{2006}]{Gal} Gallimore J. F., Axon D. J., O'Dea C. P., Baum S. A., Pedlar A. 2006, AJ, \textbf{132}, Is.2, 546-569.
\bibitem[\protect\citeauthoryear{Gallo et al.}{2015}]{Gallo} Gallo L.C., Wilkins D. R., Bonson K., et al., 2015, \mnras, \textbf{446}, 633.
\bibitem[\protect\citeauthoryear{Garc\'ia et al.}{2014}]{Garcia} Garc\'ia J. A., Dauser T., Lohfink A., et al., 2014, \apj, \textbf{782}, 76.
\bibitem[\protect\citeauthoryear{Garc\'ia, Fabian \& Kalmann}{2016}]{GFK} Garc\'ia J. A., Fabian A. C., Kalmann T. R., 2016, \mnras, \textbf{462}, 751.
\bibitem[\protect\citeauthoryear{Garc\'ia \& Kallman}{2010}]{GK} Garc\'ia J. A., Kallman T. R., 2010, \apj, \textbf{718},695.
\bibitem[\protect\citeauthoryear{Garc\'ia et al.}{2011}]{GKM} Garc\'ia J. A., Kallman T. R., Mushotzky R. F. 2011, \apj, \textbf{731},131.
\bibitem[\protect\citeauthoryear{Garc\'ia et al.}{2013}]{GDR} Garc\'ia J. A., Dauser T., Reynolds C. S. 2013, \apj, \textbf{768},146.
\bibitem[\protect\citeauthoryear{Garc\'ia et al.}{2014}]{GDL} Garc\'ia J. A., Dauser T., Lohfink A., et al. 2014, \apj, \textbf{782},76.
\bibitem[\protect\citeauthoryear{Gierli\'nski \& Done}{2004}]{GD} Gierli\'nski M., Done C., 2004, \mnras, \textbf{349}, L7.
\bibitem[\protect\citeauthoryear{Goodrich}{1989}]{G} Goodrich R. W., 1989, \apj, \textbf{342}, 224.
\bibitem[\protect\citeauthoryear{Grevesse \& Sauval}{1998}]{GS} Grevesse N., Sauval A. J., 1998, \ssr, \textbf{85},161.
\bibitem[\protect\citeauthoryear{Grier et al.}{2013}]{GMW} Grier C. J., Martini P., Watson L. C., et al. 2013, \apj, \textbf{773}, 90.
\bibitem[\protect\citeauthoryear{Grupe}{2004}]{Grupe} Grupe D., 2004, \aj, \textbf{127}, 1799.
\bibitem[\protect\citeauthoryear{Hao et al.}{2005}]{Hao} Hao C. N., Xia X. Y., Shude Mao, Hong Wu, Deng Z. G. 2005, \apj, \textbf{625}, 78.
\bibitem[\protect\citeauthoryear{J{\"a}rvel{\"a} et al.}{2017}]{JAV} J{\"a}rvel{\"a} E., L{\"a}hteenm{\"a}ki A., Lietzen H., et al., 2017, \aap, \textbf{606}, 9.
\bibitem[\protect\citeauthoryear{Jin et al.}{2012a}]{Ja} Jin C., Ward M., Done C. 2012c, \mnras, \textbf{425}, 907
\bibitem[\protect\citeauthoryear{Jin et al.}{2012b}]{J} Jin C., Ward M., Done C., et al., 2012, \mnras, \textbf{420}, 1825.
\bibitem[\protect\citeauthoryear{Jin et al.}{2013}]{Jin} Jin C. et al., 2013, \mnras, \textbf{436}, 3173.
\bibitem[\protect\citeauthoryear{Jin et al.}{2016}]{J2} Jin C., Done C., Ward M. 2016, \mnras, \textbf{455}, 691.
\bibitem[\protect\citeauthoryear{Kalberla et al.}{2005}]{K} Kalberla et al. 2005, \aap, \textbf{440}, 775.
\bibitem[\protect\citeauthoryear{Kaspi et al.}{2000}]{Ka} Kaspi S., Smith P.S., Netzer H., et al., 2000, \apj, \textbf{533}, 631.
\bibitem[\protect\citeauthoryear{Keck et al.}{2015}]{Keck} Keck M. L. et al. 2015, \apj, \textbf{806},149.
\bibitem[\protect\citeauthoryear{Laha et al.}{2013}]{Lah} Laha S. et al. 2013, \apj, \textbf{777}, 2.
\bibitem[\protect\citeauthoryear{Landi et al.}{2010}]{L}Landi R. et al., 2010, \mnras, \textbf{403}, 945.
\bibitem[\protect\citeauthoryear{Leighly}{1999}]{L1999} Leighly K. M., 1999, \apj Suppl. Ser.,\textbf{125},is.2, 317.
\bibitem[\protect\citeauthoryear{Magdziarz et al.}{1998}]{M} Magdziarz P. et al., 1998, \mnras, \textbf{301}, 179.
\bibitem[\protect\citeauthoryear{Magdziarz \& Zdziarski}{1995}]{MZ} Magdziarz P., Zdziarski A. 1995, \mnras, \textbf{273}, 837.
\bibitem[\protect\citeauthoryear{Malizia et al.}{2008}]{M1} Malizia A. et al., 2008, \mnras, \textbf{389}, 1360.
\bibitem[\protect\citeauthoryear{Malzac \& Celotti}{2002}]{MC02} Malzac J., Celotti A. 2002, \mnras, \textbf{335}, 23.
\bibitem[\protect\citeauthoryear{Maraschi \& Haardt}{1997}]{MH} Maraschi L., Haardt F. 1997, in IAU Colloquium 163, ASP Conference Series, ed. D. T. Wickramasinghe, G. V. Bicknell and L. Ferrario, \textbf{121}, 101.
\bibitem[\protect\citeauthoryear{Marconi et al.}{2006}]{M2} Marconi A. et al. 2006, \apj, \textbf{678}, 693.
\bibitem[\protect\citeauthoryear{Mathur}{2011}]{MAT} Mathur S., 2011, in Proc. of the Workshop Narrow-Line Seyfert 1 Galaxies
and Their Place in the Universe, ``Host galaxies of NLS1s'', eds. L. Foschini, M. Colpi, L. Gallo, D. Grupe, S. Komossa, K. Leighly \& S. Mathur, Milano, Italy, April 4-6, 2011, p. 35.
\bibitem[\protect\citeauthoryear{Matt et al.}{2015}]{MATT} Matt G., Balokovic\' A., Marinucci A. et al., 2015, \mnras, \textbf{447}, 3029.
\bibitem[\protect\citeauthoryear{Maza \& Ruiz}{1989}]{MR} Maza J., Ruiz M.T. 1989, \apjs, \textbf{69}, 353.
\bibitem[\protect\citeauthoryear{Merloni et al.}{2006}]{MERL} Merloni A. et al., 2006, \mnras,\textbf{370}, 1699.
\bibitem[\protect\citeauthoryear{Mehdipour et al.}{2011}]{MEH} Mehdipour M. et al., 2011, \aap, \textbf{444}, 1469.
\bibitem[\protect\citeauthoryear{Miniutti et al.}{1999}]{M3} Miniutti G., 2009, \mnras, \textbf{398}, 255.
\bibitem[\protect\citeauthoryear{Morrison \& MackCammon}{1983}]{MM} Morrison R., McCammon D., 1983, \apj, \textbf{270}, 119.
\bibitem[\protect\citeauthoryear{Mushotzky et al.}{1993}]{MDP} Mushotzky R., Done C., Pounds K.A. 1993, \araa, \textbf{31}, 717.
\bibitem[\protect\citeauthoryear{Nandra \& Pounds}{1994}]{NP} Nandra K., Pounds K.A., 1994, \mnras, \textbf{268}, 405.
\bibitem[\protect\citeauthoryear{Nandra et al.}{2007}]{N} Nandra K., O\'Neill P. M., George I.M., Reeves J. N. 2007,\mnras, \textbf{382}, 194.
\bibitem[\protect\citeauthoryear{Nayakshin et al.}{2000}]{NKK} Nayakshin S., Kazanas D., Kallman T. R. 2000, \apj, \textbf{537}, 833.
\bibitem[\protect\citeauthoryear{Orlandini et al.}{2012}]{Orl} Orlandini et al., 2012, \apj, \textbf{748}, 86.
\bibitem[\protect\citeauthoryear{Osterbrock \& De Robertis}{1985}]{OdR} Osterbrock D. E., De Robertis M. M., 1985, \pasp, \textbf{97}, 1129.
\bibitem[\protect\citeauthoryear{Osterbrock \& Pogge}{1985}]{OP} Osterbrock D. E., Pogge R. 1985, \apj, \textbf{297}, 166.
\bibitem[\protect\citeauthoryear{Pal et al.}{2016}]{P} Pal M. et al., 2016, \mnras, \textbf{457}, 875.
\bibitem[\protect\citeauthoryear{Panessa et al.}{2011}]{PdRB} Panessa F., De Rosa A., Bassani L., et al., 2011, \mnras, \textbf{417}, 2426.
\bibitem[\protect\citeauthoryear{Parker et al.}{2014}]{Par} Parker M.L. et al., 2014, \mnras, \textbf{443},1723.
\bibitem[\protect\citeauthoryear{Paturel et al.}{2002}]{PDBW} Paturel G., Dubois P., Petit C., Woelfel F. 2002, LEDA.
\bibitem[\protect\citeauthoryear{Pfefferkorn, Boller \& Rafanelli}{2001}]{PBR} Pfefferkorn F., Boller T., Rafanelli P. 2001, \aap, \textbf{368}, 797.
\bibitem[\protect\citeauthoryear{Pounds, Done \& Osborne}{1995}]{PDO} Pounds K. A., Done C., Osborne J. P. 1995, \mnras, \textbf{277}, L5.
\bibitem[\protect\citeauthoryear{Protassov et al.}{2002}]{Prot} Protassov R. et al., 2002, \apj, \textbf{571}, 545.
\bibitem[\protect\citeauthoryear{Pulatova et al.}{2015}]{Pul} Pulatova N.G. et al., 2015, \mnras, \textbf{447}, 2209.
\bibitem[\protect\citeauthoryear{Reynolds et al.}{2012}]{Rey}Reynolds C. R., Brenneman L. W., Lohfink A. M., et al. 2012, \apj, \textbf{755}, 88.
\bibitem[\protect\citeauthoryear{Ricci et al.}{2011}]{RWCP} Ricci C., Walter R., Courvoisier T. J.-L., Paltani S., 2011, \aap, \textbf{532}, A102.
\bibitem[\protect\citeauthoryear{Sambruna et al.}{2011}]{Sam} Sambruna R. M. et al. 2011, \apj, \textbf{734},105.
\bibitem[\protect\citeauthoryear{Schlafty \& Finkbeiner}{2011}]{SF} Schlafty E. F., Finkbeiner D. P., 2011, \apj, \textbf{737}, 103.
\bibitem[\protect\citeauthoryear{Schlegel, Finkbeiner \& Davis}{1998}]{SFD} Schlegel D. J., Finkbeiner D. P., Davis M., 1998, \apj, \textbf{500}, 525.
\bibitem[\protect\citeauthoryear{Sobolewska \& Done}{2005}]{SD}Sobolewska M., Done C., 2005, in AIP Conf. Proc. \textbf{774}, X-Ray Diagnostics of Astrophysical Plasmas: Theory, Experiment, and Observation, ed. R. K. Smith (Melville: AIP), 317.
\bibitem[\protect\citeauthoryear{Starling et al.}{2013}]{Star} Starling R. L. C., Done C., Jin C. et al. 2014, \mnras, \textbf{437}, 3929.
\bibitem[\protect\citeauthoryear{Sulentic et al.}{2000}]{S} Sulentic J. W. et al., 2000, \apj, \textbf{536}, L5.
\bibitem[\protect\citeauthoryear{Sulentic et al.}{2008}]{S1} Sulentic J. W. et al., 2008, revMexAA(SC), \textbf{32}, 51.
\bibitem[\protect\citeauthoryear{Terashima et al.}{2009}]{T} Terashima Y. et al., 2009, \pasj, \textbf{61}, 299.
\bibitem[\protect\citeauthoryear{Valencia et al.}{2012}]{VZE} Valencia S. M., Zuther J., Eckart A. et al. 2012, Proc. Nuclei of Seyfert galaxies and QSOs - Central engine \& conditions of star formation, Bonn, (arXiv:1305.3273).
\bibitem[\protect\citeauthoryear{Vasylenko et al.}{2015}]{VZF} Vasylenko A. A., Zhdanov V. I., Fedorova E. V., 2015, \apss, \textbf{360}, 16.
\bibitem[\protect\citeauthoryear{V\'eron-Cetty \& V\'eron}{2006}]{VV} V\'eron-Cetty M.-P., V\'eron P., 2006, \aap, \textbf{455}, 773.
\bibitem[\protect\citeauthoryear{V\'eron-Cetty et al.}{2001}]{VVG} V\'eron-Cetty M.-P., V\'eron P., Gonsales A. C. 2001, \aap, \textbf{372}, 730.
\bibitem[\protect\citeauthoryear{Wang \& Lu}{2001}]{WL} Wang T., Lu Y. 2001, \aap, \textbf{377}, 52.
\bibitem[\protect\citeauthoryear{Wilms et al.}{2000}]{WAM} Wilms J., Allen A., McCray R. 2000, \apj, \textbf{542}, 914.
\bibitem[\protect\citeauthoryear{Young \& Wilson}{2004}]{YW}Young  A. J.,Wilson  A. S. 2004, \apj, \textbf{601}, 133.
\bibitem[\protect\citeauthoryear{Zhou et al.}{2006}]{Z} Zhou, H.-Y. et al. 2006, \apjs, \textbf{166}, 128.
\bibitem[\protect\citeauthoryear{Zoghbi et al.}{2010}]{Z1} Zoghbi A. et al. 2010, \mnras, \textbf{401}, 24.

\end{thebibliography}

%

\begin{table*}
\centering
\begin{threeparttable}
\caption{\fontsize{9}{13}\selectfont Best-fit parameters to the EPIC/PN spectrum for baseline models.}\label{tab1}
\fontsize{8}{12}\selectfont
\begin{tabular}{|c|c|c|c|c|} 
\hline \multicolumn{2}{|c|}{Continuum model:} & {\tt zpowerlw} & {\tt pexmon} & {\tt xillver} \\\hline
\multicolumn{2}{|c|}{$\chi^2$/d.o.f.} & 426/423 & 419/426 & 420/426 \\\hline \hline
Parameter & Unit & \multicolumn{3}{|c|}{Value}\\\hline
kT & eV & 148$\pm$0.01 & 150$\pm$0.03 & 155$\pm$0.04 \\
$\tau$ & & 0.23$\pm$0.11 & 0.23$\pm$0.06 & 0.23$\pm$0.08 \\
E$_{1 edge}$ & keV & 0.60$\pm$0.02 & 0.60$\pm$0.01 & 0.60$\pm$0.01 \\
$\tau_{1 edge}$ & & 0.67$_{-0.34}^{+0.50}$ & 0.65$_{-0.19}^{+0.23}$ & 0.79$_{-0.18}^{+0.21}$ \\
E$_{2 edge}$ & keV & 0.73$\pm$0.04 & 0.73$\pm$0.02 & 0.74$\pm$0.02 \\
$\tau_{2 edge}$ & & 0.67$_{-0.22}^{+0.28}$ & 0.71$_{-0.12}^{+0.14}$  & 0.66$_{-0.13}^{+0.15}$  \\
$\Gamma$ & & 2.11$\pm$0.07 & 2.12$\pm$0.04 & 2.09$\pm$0.04 \\
$R$ & & - & 0.84$\pm$0.36 & - \\
$\xi$ & $erg~s^{-1}~cm$ & - & - & 61$_{-27}^{+63}$\\
Fe$_{abund}$ & & - & 1.16$_{-0.86}^{+1.16}$ & 13.23$_{-1.73}^{+1.49}$ \\
E$_{FeK}$ & keV & 6.43$_{-0.27}^{+0.07}$ & - & - \\
$\sigma_{FeK}$ & eV & 10(fixed) & - & - \\
E$_{FeXXVI}$ & keV & 7.03$_{-0.33}^{+0.19}$ & 7.06$_{-0.49}^{+0.22}$ & - \\
$\sigma_{FeXXVI}$ & eV & 235$_{-126}^{+77}$& 272$_{-183}^{+1009}$ & - \\\hline
\multicolumn{2}{|c|}{$L^{intr}_{0.5-2.0 keV}$} & 4.74$\pm$0.03 & 4.73$_{-0.03}^{+0.04}$ & 4.74$_{-0.04}^{+0.03}$ \\
\multicolumn{2}{|c|}{$L^{intr}_{2.0-10.0 keV}$} & 4.17$_{-0.09}^{+0.10}$ & 4.25$\pm$0.09 & 4.19$\pm$0.08 \\\hline

\end{tabular}
    \begin{tablenotes}
     \item[1] Luminosities are in $10^{42}$ erg s$^{-1}$ units and are corrected for absorption.
   \end{tablenotes}
\end{threeparttable}
\vspace{0.05in}

\vskip20pt
\centering
\begin{threeparttable}
 \caption{\fontsize{9}{13}\selectfont Best-fit parameters to the broad-band data for two different reflection models.}\label{tab2}
 \fontsize{8}{12}\selectfont
\begin{tabular}{|c|c|c|} 
\hline Data/ & {PN+BAT} & {PN+ISGRI} \\
parameter & & \\
\hline \hline
\multicolumn{3}{|c|}{continuum model: \tt pexmon}\\
\hline
$\chi^2$/d.o.f. & 426/434 & 425/438 \\
$\Gamma$ & 2.13$\pm$0.04 & 2.14$\pm$0.03 \\
$R$ & 0.91$\pm$0.33 & 0.91$\pm$0.35 \\
Fe$_{abund}$ & 1(fixed) & 1(fixed) \\
\hline
$L^{intr}_{0.5-2.0 keV}$ & 4.74$\pm$0.03 & 4.74$_{-0.03}^{+0.04}$ \\
$L^{intr}_{2.0-10.0 keV}$ & 4.17$_{-0.07}^{+0.10}$ & 4.17$_{-0.08}^{+0.10}$ \\
$L^{intr}_{20-100 keV}$ & 4.01$_{-0.16}^{+1.29}$ & 3.83$_{-0.26}^{+0.06}$ \\\hline
\multicolumn{3}{|c|}{continuum model: \tt xillver}\\
\hline
$\chi^2$/d.o.f. & 427/434 & 426/438 \\
$\Gamma$ & 2.08$\pm$0.02 & 2.09$\pm$0.04 \\
$\xi$, $erg~s^{-1}~cm$ & 62$_{-21}^{+61}$ & 62$_{-28}^{+69}$ \\
Fe$_{abund}$ & 3.19$_{-1.70}^{+1.28}$ & 3.19$_{-1.73}^{+1.48}$ \\
\hline
$L^{intr}_{0.5-2.0 keV}$ & 4.73$\pm$0.04 & 4.73$_{-0.03}^{+0.05}$ \\
$L^{intr}_{2.0-10.0 keV}$ & 4.23$\pm$0.09 & 4.24$\pm$0.08 \\
$L^{intr}_{20-100 keV}$ & 4.12$_{-0.98}^{+1.09}$ & 3.97$_{-0.35}^{+0.10}$\\\hline
\end{tabular}
 \begin{tablenotes}
     \item[1] The luminosity units are the same as used in Table~\ref{tab1}.
   \end{tablenotes}
\vspace{0.05in}
\end{threeparttable}
\end{table*}

\vskip20pt
\begin{table*}
\centering
\begin{threeparttable}
\caption{\fontsize{9}{13}\selectfont Best-fit parameters for the inrinsic disc comptonized model to the UV+EPIC/PN data}\label{tab3}
\fontsize{8}{12}\selectfont
\begin{tabular}{|c|c|c|c|}
\hline Model component & Parameter & Unit & Value \\\hline \hline
\multirow{2}{*}{\tt xillver} & $\xi$ & $erg~s^{-1}~cm$ & 21$_{-12}^{+31}$ \\
              & Fe$_{abund}$ & & 3.11$_{-2.01}^{+1.61}$ \\\hline
\multirow{7}{*}{\tt optxagnf} & $\Gamma$ & & 1.94$\pm$0.01 \\
              & kT & eV & 487 $\pm$0.07 \\
              & $\tau$ & & 8.82$_{-0.94}^{+0.89}$ \\
              & $M_{SMBH}$ & 10$^6$ M$_{\odot}$ & 6.9$\pm$0.4 \\
              & $log_{L/L_{Edd}}$ & & $-0.57\pm0.02$ \\
              & R$_{cor}$ & $R_{g}$ & $11.3\pm0.5$ \\
              & $f_{pl}$ & & $0.53\pm0.01$\\\hline
\multicolumn{2}{|c|}{$\chi^2$/d.o.f.} & \multicolumn{2}{|c|}{426/428}\\
\multicolumn{2}{|c|}{$L^{intr}_{0.5-2.0 keV}$} & \multicolumn{2}{c|}{4.81$\pm$0.04} \\
\multicolumn{2}{|c|}{$L^{intr}_{2.0-10.0 keV}$} & \multicolumn{2}{c|}{4.07$_{-0.09}^{+0.07}$} \\\hline
\end{tabular}
 \begin{tablenotes}
     \item[1] The luminosity units are the same as used in Table~\ref{tab1}.
   \end{tablenotes}
\end{threeparttable}
\vspace{0.05in}
\end{table*}

\vskip20pt
\begin{table*}[!tp]
\centering
\begin{threeparttable}
\caption{\fontsize{9}{13}\selectfont Best-fit parameters for two ionized reflection model obtained from the three broad-band datasets}\label{tab4}
\fontsize{8}{12}\selectfont
\begin{tabular}{|c|c|c|c|c|c|}
\hline
\multicolumn{3}{|c|}{Data:} & PN+BAT & PN+ISGRI & PN+ISGRI+BAT \\\hline
\multicolumn{3}{|c|}{$\chi^2$/d.o.f.} & 425/433 & 418/437 & 429/443 \\\hline
Model component & Parameter & Unit & \multicolumn{3}{|c|}{Value} \\\hline\hline
{\tt zbremss} & kT & eV & 296$\pm36$ & 309$\pm30$ & 299$\pm18$ \\
{\tt zedge} & E$_{edge}$ & keV & 0.592$\pm0.04$ & 0.613$\pm0.03$ & 0.619$\pm0.024$ \\
            & $\tau$ & & 0.20$_{-0.13}^{+0.18}$ & 0.39$_{-0.18}^{+0.30}$ & 0.41$\pm0.10$ \\
{\tt zpowerlaw} & $\Gamma$ & & 2.10$\pm0.06$ & 2.07$\pm0.05$ & 2.06$\pm0.05$ \\\hline
\multicolumn{6}{|c|}{mildly ionized reflection} \\\hline
\multirow{3}{*}{\tt xillver} & $\Gamma$ & & 2.10(tied) & 2.07(tied) & 2.06(tied) \\
              & $\xi$ & $erg~s^{-1}~cm$ & 30$_{-18}^{+30}$ & 37$_{-17}^{+39}$ & 51$_{-25}^{+65}$ \\
              & Fe$_{abund}$ & & 4.46$_{-0.94}^{+1.28}$ & 4.34$_{-1.27}^{+3.13}$ & 4.86$_{-1.41}^{+4.48}$ \\\hline
\multicolumn{6}{|c|}{highly ionized reflection} \\\hline
\multirow{3}{*}{\tt xillver} & $\Gamma$ & & 2.10(tied) & 2.07(tied) & 2.06(tied) \\
              & $\xi$ & $erg~s^{-1}~cm$ & 2731$_{-410}^{+988}$ & 3070$_{-1234}^{+3293}$ & 3838$_{-1612}^{+3160}$ \\
              & Fe$_{abund}$ & & 4.46(tied) & 4.34(tied) & 4.86(tied) \\
\hline
\multicolumn{3}{|c|}{$L^{intr}_{0.5-2.0 keV}$} & 4.78$\pm$0.04 & 4.76$\pm$0.03 & 4.75$\pm$0.03 \\
\multicolumn{3}{|c|}{$L^{intr}_{2.0-10.0 keV}$} & 4.02$\pm$0.10 & 4.08$\pm$0.10 & 4.04$\pm$0.01 \\
\multicolumn{3}{|c|}{$L^{intr}_{20-100 keV}$} & 4.03$_{-0.91}^{+1.18}$ & 3.78$_{-0.34}^{+0.01}$ & 3.72$_{-0.29}^{+0.03}$ \\\hline

\end{tabular}
 \begin{tablenotes}
     \item[1] The luminosity units are the same as used in Table~\ref{tab1}.
   \end{tablenotes}
\end{threeparttable}
\vspace{0.05in}
\end{table*}
\end{document}